\documentclass[twocolumn,nofootinbib]{revtex4-1}
\usepackage{latexsym,epsfig,amssymb, amsmath,nicefrac}

\usepackage{amsfonts,amsthm} 
\usepackage{bm,bbm}
\usepackage{graphicx}
\usepackage{esint}
\usepackage{color}
\usepackage{braket}

\newcommand{\be}{\begin{equation}}
\newcommand{\ee}{\end{equation}}
\newcommand{\ba}{\begin{eqnarray}}
\newcommand{\ea}{\end{eqnarray}}
\def\bs{\begin{subequations}}
\def\es{\end{subequations}}

\newcommand{\rf}[1]{(\ref{#1})}

\begin{document}


\title{{\Large The double attractor behavior of  induced inflation}}

\author{Renata Kallosh${}^1$, Andrei Linde${}^1$ and Diederik Roest${}^2$}

\affiliation{{}$^1$Department of Physics and SITP, Stanford University, \\ 
Stanford, California 94305 USA, kallosh@stanford.edu, alinde@stanford.edu}

\affiliation{{}$^2$Centre for Theoretical Physics, University of Groningen, \\ Nijenborgh 4, 9747 AG Groningen, The Netherlands, d.roest@rug.nl}

\begin{abstract}
We describe  an induced inflation, which refers to a class of inflationary models with a generalized non-minimal coupling $\xi g(\phi) R$ and a specific scalar potential. The defining property of these models is that the scalar field takes a vev in the vacuum and thus induces an effective Planck mass. We study this model as a function of the coupling parameter $\xi$. At large $\xi$, the predictions of the theory are known to have an attractor behavior, converging to a universal result independent on the choice of the function $g(\phi)$. We find that at small $\xi$, the theory approaches a second attractor. The inflationary predictions of this class of theories continuously interpolate between those of the Starobinsky model and the predictions of the simplest chaotic  inflation with a quadratic potential.\end{abstract}

\maketitle

\smallskip

\noindent {\bf Introduction.\ } 
Inflationary models are specified not only by the inflaton's kinetic term and potential energy, but also its coupling to the gravitational sector. This (non-)minimal coupling {is a natural ingredient of effective field theories involving gravity and a scalar,} and can be parametrized by $\sqrt{-g}  \, \tfrac12 \Omega (\phi) R_J$;  this is referred to as the Jordan frame if $\Omega (\phi)$ is a non-trivial function of scalars. The cosmological observables are extremely sensitive to even small changes in this non-minimal coupling, which therefore deserves close study. 

A case in point is the  $\lambda\phi^{4}$ model with {the often considered $\Omega(\phi) = 1 + \xi\phi^{2}$, which has a conformal symmetry for the special case $\xi = -1/6$. We are interested in positive values, however:} while $\xi=0$ is ruled out by Planck data  \cite{Ade:2013uln}, the same model {with $\xi \gtrsim 2\cdot 10^{-3}$} is already quite consistent with observations  \cite{Bezrukov:2013fca}. At large $\xi$ predictions of this particular model tends to a low $r$ Higgs inflation model \cite{Salopek:1988qh}. The inflaton potential for this model in the Einstein frame at large  $\xi$ and large absolute values of the canonically normalized inflaton field $\varphi$ acquires the form
\be\label{H}
V_{H} \sim (1 - e^{-\sqrt{2  /3}\, |\varphi |})^2\ ,
\ee
up to an overall numerical coefficient which depends on $\lambda$ and $\xi$. 
Interestingly, for $\varphi> 0$ this potential coincides with the potential of the Starobinsky model $R+R^{2}$ \cite{Starobinsky:1980te} in the representation proposed by Whitt \cite{Whitt:1984pd}, 
\be\label{S}
V_{S} \sim (1 - e^{-\sqrt{2  /3}\, \varphi })^2\ .
\ee
In both models, inflation  leads to the universal prediction 
 \be
n_{s} =1-2/N\, , \qquad r = 12/N^{2} 
\label{attractor}\ee
in the leading approximation in $1/N$, where $N$ is the number of e-folding of inflation. Note, however, that the shape of the potentials   at negative $\varphi$ is very different in these two models. In Higgs inflation, the potential is even in $\varphi$, and inflation can occur both at $\varphi > 0$ and at $\varphi < 0$. Meanwhile, in the Starobinsky model, the potential at large negative $\varphi$ blows up exponentially, and inflation is possible only at $\varphi > 0$.

During the last year, several different classes of models with a similar attractor behavior have been found and their supergravity generalizations have been constructed. They include, in particular, a broad class of models of conformal  inflation \cite{Kallosh:2013hoa},  $\alpha$-attractors \cite{Kallosh:2013yoa}, and the models with generalized non-minimal coupling \cite{Kallosh:2013tua}; see \cite{Linde:2014nna} for a recent review.

In particular, it was shown in \cite{Kallosh:2013tua} that this attractor behavior is valid for a broad class of functions 
$\Omega (\phi)= 1 +\xi f(\phi)$
 and Jordan frame potentials $V_J=\lambda f^2(\phi)$. {This can be seen as a natural generalization of the special case $f(\phi) = \phi^2$ considered before.} Predictions of all of these  models at large  $\xi$ asymptote to the universal attractor point (\ref{attractor}). Similar results have been later found for induced inflation models with $\Omega (\phi)= \xi g(\phi)$  \cite{Giudice:2014toa,Kallosh:2014rha}.

 For many of such models, just like for the model $\lambda\phi^{4}$, the convergence to the strong coupling limit begins very early, already at $\xi \ll 1$, so the speculations about the unitarity bound in these theories for $\xi \sim  10^{4}$ are largely irrelevant for the discussion of the attractor behavior found in  \cite{Kallosh:2013tua}; see  a detailed discussion of these issues in  \cite{Kallosh:2013tua,Linde:2014nna, Edwards}. 
 
The recent BICEP2 data release \cite{Ade:2014xna} has attracted attention to the possibility that models with non-minimal coupling may be consistent with observations not only in the strong coupling limit, but in the weak coupling limit as well, for $\xi \to 0$, far from the universal attractor. According to  \cite{Kallosh:2013tua}, for the theories with $\Omega (\phi)= 1 +\xi f(\phi)$, the limit $\xi\to 0$ simply restores  {\it different} model-dependent predictions of  various chaotic inflation models with minimal coupling to gravity \cite{Linde:1983gd}.
However, as we will show in this paper, the predictions of the induced inflation $\Omega (\phi)= \xi g(\phi)$ in the limit $\xi \to 0$ converge to the single {\it universal} prediction, the one of the simplest version of chaotic inflation with the quadratic potential,
\be \label{q}
n_{s} =1-2/N\, , \qquad r = 1/8N \ .
\ee
Interestingly, this class of models therefore interpolates between two different inflationary attractors.

\smallskip

\noindent {\bf Universal attractor.\ } 
We will first review the universal attractor that arises for a sufficiently strong non-minimal coupling $\xi\rightarrow \infty $ of a particular form, following \cite{Kallosh:2013tua,Giudice:2014toa,Kallosh:2014rha}. The starting point is the most general single-field and two-derivative inflationary action: it is defined in terms of three functions $\Omega (\phi), K(\phi), U(\phi)$ and reads
\be
{\cal L}_J = \sqrt{-g} \left[\tfrac12 \Omega(\phi) R -  \tfrac12 K(\phi) (\partial \phi )^2 -  U(\phi) \right] \, .
\label{lagJ}
\ee
The above will be referred to as the Jordan frame formulation. For $\Omega(\phi)  > 0$, by means of a conformal transformation,  one can always go to the Einstein frame, corresponding to setting $\Omega = 1$. This implies the following for the kinetic terms and potential energy of the theory:
  \be
{\cal L}_E=\sqrt{-g} \left[ \frac{R}{2}  -\frac{1}{2} \left( \frac{K}{\Omega}+\frac{3{\Omega^\prime}^2}{2\Omega^2}\right) (\partial \phi )^2 -\frac{U}{\Omega^2} \right] \, .
\label{lagE}
\ee
We will further restrict our attention to the particular case
 \begin{align}\label{universal}
  K = 1 \,, \quad U = \lambda (\Omega - 1)^2 \,, \quad \Omega (\phi)= 1 +\xi f(\phi) \ ,
 \end{align}
with the function $f$ vanishing at the origin \cite{Kallosh:2013tua}.

The universal attractor at $\xi\rightarrow \infty $ arises when the second part of the kinetic term dominates over the first:
 \begin{align}
   \frac{3{\Omega^\prime}^2}{2\Omega^2}\gg {1\over \Omega} \,. \label{kin-approx}
 \end{align} 
In this situation, one can define a canonically normalized  scalar field
 \begin{align}
   \varphi = \pm \sqrt{\tfrac32} \log \Omega(\phi) \,. \label{canonical}
 \end{align}
The choice of sign here is non-trivial because $\log \Omega(\phi)$ vanishes at $\phi = 0$, so one should be careful to make sure that the function $\varphi(\phi)$ is differentiable at $\phi = 0$. If $f (\phi)$ is an odd function of $\phi$, e.g.~linear, the simple expression \eqref{canonical} with a definite sign remains a solution for $-\infty < \phi < \infty$. In contrast, if  $f (\phi)$ is even, e.g.~quadratic, then one should use both solutions with opposite signs for $\phi < 0$ and $\phi > 0$.
 
As a result,  if $f (\phi)$ is an odd function of $\phi$,  the Einstein frame inflaton potential in the  large $\xi$ limit coincides with the potential in the Starobinsky model (\ref{S}). Markedly different, for even  $f (\phi)$, it coincides with the large $\xi$ limit of the Higgs inflation potential (\ref{H}) \cite{Kallosh:2013tua}.
Note that these potentials differ for negative $\varphi$,  but they have an identical inflationary plateau for positive $\varphi$. As inflation only takes place at a single plateau, the predictions of these potentials are identical and are given by \rf{attractor}. 

The situation is even simpler in the Higgs model when the field in the unitary gauge is represented  by its radial degree of freedom taking only positive values. There the difference between these two potentials disappears for all values of the Higgs field.

This set of models allow various generalizations described in  \cite{Kallosh:2013tua,Giudice:2014toa,Kallosh:2014rha}. Here we will discuss one of them, called induced inflation \cite{Giudice:2014toa,Kallosh:2014rha}.

\smallskip

\noindent {\bf Induced inflation.\ } 
In the induced inflation model one takes
 \begin{align}
  \Omega = \xi g(\phi) \,, \qquad U = \lambda \bigl(g(\phi) - \xi^{-1}\bigr)^2 \, ,  \label{induced}
 \end{align}
with $g(0)= 0$. The coefficient of the Einstein-Hilbert term is determined by the vev of the scalar field: as the scalar potential has a minimum at $\Omega = 1$, this gives rise to the usual expression for the Planck mass.

Note that  the models \eqref{universal} and \eqref{induced} are related by the simple replacement $f(\phi) \rightarrow g(\phi) - \xi^{-1}$ (for more details and the embedding in supergravity, see \cite{Kallosh:2014rha, Pallis}). Their strong coupling limits for large $g(\phi)$ are therefore similar. However,   the detailed structure of the inflaton potential in these theories is  different. 

As an example, let us consider the simplest induced inflation model with $\Omega = \xi\phi^{2}$ at sufficiently large $\xi$. In this case, the approximation \eqref{kin-approx} is valid for all $\phi >0$. Therefore one can use equation (\ref{canonical}) for all $\phi$, which yields
\be
\varphi = \pm  \sqrt{\tfrac32} \log \, (\xi\phi^{2})  \,.  \label{canonical-induced}
\ee
In this theory, just like in the theory with $\Omega = 1+\xi\phi^{2}$, one should be careful when analyzing the solution at $\phi = 0$, but for a different reason. The point $\phi= 0$ is the boundary of the moduli space for induced inflation, which corresponds to infinitely large values of $|\varphi|$. In the Jordan frame it corresponds to the singularity of the effective gravitational constant, whereas in the Einstein frame the potential diverges in the limit $\phi \to 0$. As a results, two branches of the solution \eqref{canonical-induced} are disconnected, and only one of them describes our world. Without loss of generality, one can use the positive branch, which relates all physically accessible values $0<\phi<\infty$ to the full range of values of the canonically normalized inflaton field, $-\infty< \varphi < \infty$. This conclusion remains valid for the Higgs field taking values $0<\phi<\infty$ in the unitary gauge, but the full range of values of the canonically normalized inflaton field, $-\infty< \varphi < \infty$.

As a result, induced inflation at strong coupling has a canonical scalar field defined by 
\be
\Omega =e^{\sqrt{2/3 }\, \varphi}, \ee with $-\infty< \varphi < \infty$. 
Hence the potential in this model is given by the Starobinsky-Whitt equation (\ref{S}) rather than by the Higgs inflation type expression (\ref{H}) given in \cite{Giudice:2014toa}. The difference disappears for inflation at $\varphi >0$, but it  leads to the absence of induced inflation at large $\xi$ for $\varphi < 0$. 

Finally, one should note that in other, more complicated versions of induced inflation, the potential at large $\xi$ is also given by the expression (\ref{S}) in the regions where the condition  \rf{kin-approx} is satisfied, but under certain conditions inflation may take place even when this condition is violated and the expression for the potential becomes somewhat different from (\ref{S}).

\smallskip

\noindent {\bf Quadratic attractor.\ } 
Turning to the opposite limit of weak coupling, $\xi \rightarrow 0$, the differences between the universal attractor with \rf{universal} and induced inflation \rf{induced}  increase. Indeed, as we will see, their predictions diverge wildly in this limit: while the non-minimal coupling  \rf{universal} leads to the minimally coupled original model in this limit, the induced inflation models \rf{induced} typically lead to the same observational consequences as the simplest chaotic inflation model with a quadratic potential.

More concretely, in this limit, we will be interested in the case where the first part dominates the kinetic term in \eqref{lagE}, exactly opposite to requirement for the universal attractor (see \cite{Mosk} for a related discussion). Inflation will take place close to the minimum $\Omega =1$, and hence requires very large values of $g$ as $\xi$ becomes small. Moreover,  $g'/ g$ is required to be negligible during inflation. This leads to the approximation
 \begin{align}
  g = \xi^{-1} + g_1 \varphi \,.
 \label{expansion}
 \end{align}
The above condition requires the first Taylor coefficient $g_1$ to be much smaller than the zeroeth coefficient $\xi^{-1}$: we are expanding the function $g$ at very large field values, but require its derivative to be smaller.  Therefore, to lowest order in $\xi g_1 < 1$, the kinetic terms are canonical in this limit. Moreover, the scalar potential becomes 
 \begin{align}
  V = \lambda g_1^2 \varphi^2 \,, 
 \end{align}
again at lowest order in $\xi g_1$, and therefore coincides with quadratic inflation in the limit $\xi \rightarrow 0$. The corresponding inflationary predictions are therefore given by (\ref{q}) at lowest order in $1/N$.

We have thus found that this model has two different attractors in these opposite limits. The requirements for the Starobinsky attractor at strong coupling is a zero of the function $g$ while $g'/g$ becomes large. In contrast, the quadratic attractor arises for functions $g$ that can become arbitrarily large at some field value, while $g'/ g$ is smaller than one.  Interestingly, both attractors have identical spectral indices at lowest order, but differ in their expressions for the tensor-to-scalar ratio. These two possibilities were identified as the two universality classes with this value of $n_s$ \cite{Mukhanov:2013tua,Roest:2013fha}.

One should note that the existence of the double attractor regime in the induced inflation  is very similar to
the recently revealed property of $\alpha$-attractors \cite{Kallosh:2014rga}, where it was shown that in the limit $\alpha \to  0$ the predictions of a broad class of $\alpha$-attractors  \cite{Kallosh:2013yoa} converge to  $n_{s} =1-2/N,$   $r = 0$, whereas in the opposite limit $\alpha \to \infty$ the predictions converge to those of the chaotic inflation with a quadratic potential \rf{q}.

The quadratic attractor in the induced inflation actually arises in a more general setting than discussed above. First of all, similar to the universal attractor \cite{Kallosh:2013tua}, the two functions in \eqref{induced} are not necessarily equal; as long as they have an identical expansion \eqref{expansion} at lowest order around the value $\xi^{-1}$, the weak coupling limit will give rise to \eqref{q}. As an example, any two polynomials that have terms up to the same power have this property. This is to be contrasted with the requirement for the universal attractor, where the two functions  are required to have terms from the same power onwards. More concretely, consider the polynomials
 \begin{align}
  \Omega = \xi \sum_{i=m}^p c_i \phi^i \,, \qquad U = \lambda \Bigl(\sum_{i=n}^q d_i \phi^i - \xi^{-1}\Bigr)^2 \,,
 \end{align}
with non-vanishing coefficients $c_i$ and $d_i$. These will asympote to the universal attractor at large $\xi$ provided $m=n$, and to the quadratic attractor at small $\xi$ if $p=q$.

Finally, one can include an arbitrary smooth function $h(\Omega)$ in the scalar potential,
 \begin{align}
  U = \lambda h(\Omega) (\Omega-1)^2 \,. \label{induced1}
 \end{align}
Provided $h$ does not vanish at $\Omega=1$, such a function will not alter the weak coupling limit. However, the function $h$ will generically destroy the universal attractor in the strong coupling limit $\xi \gg 1$, while preserving the attractor at $\xi \ll 1$.

\begin{figure}[t!]
\includegraphics[scale=.8]{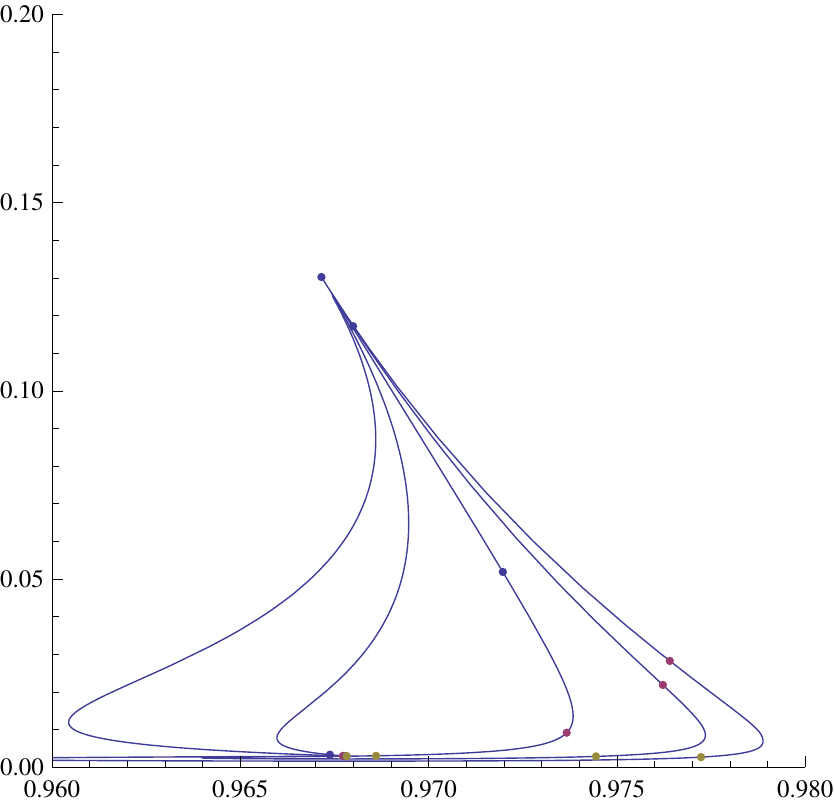}
 \includegraphics[scale=.8]{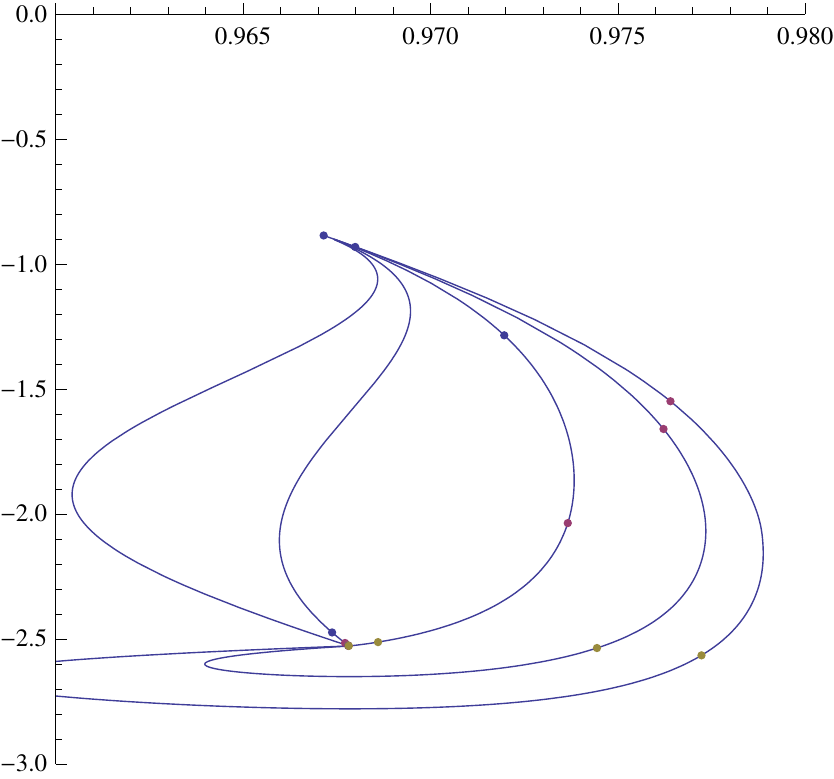}
\caption{\it \small{The $\zeta$-dependence of $(n_s, \log(r))$ for different chaotic models with $n=(2/3, 1, 2, 6, 16)$ (in decreasing redshift, i.e.~from right to left) for 60 e-folds. The dots correspond to $\log(\xi) = (-1,0,1)$.}}
\vspace{-.3cm}
\end{figure}

\smallskip

\noindent {\bf Examples.\ } 
A numerical analysis confirms the prediction that the induced inflation models interpolate between Starobinsky and quadratic inflation. As an example, we take the function $g$ defined in \eqref{induced} to be
 \begin{align}
  g =\phi^{n/2} \,.
 \end{align}
The resulting plots are given in figure 1 for various values of $n$. Note that all lines initially point in the same direction. This is a consequence of the universal behavior for small $\xi$ identified in \cite{Kallosh:2013tua}: irrespective of the  minimally coupled model, all lines initially have a slope of $-16$ in the $(n_s,r)$ plane at weak coupling. In contrast to \cite{Kallosh:2013tua}, we do not find a comb-like structure at weak coupling; this is a consequence of the quadratic attractor for induced inflation at weak coupling. In the strong coupling behaviour, the observational predictions between universal and induced inflation are negligible, and we recover the same approach to the Starobinsky-like attractor.

The transition point between the two different limiting behaviors lies around $\xi = 1$ (purple dots). More precisely, for higher $n$ one needs an increasingly small $\xi$ in order to get close to the quadratic point. We anticipate that the tipping point scales as $\log(\xi) = - (n+1)/2$ as a function of the power of the monomial. In other words, while for $n=1$ it suffices to take $\xi = 1/10$ in order to approach quadratic, we expect that for e.g.~$n=5$ a similar asymptotic regime is only reached when $\xi$ becomes of order $10^{-3}$.

A special case in the above, already considered by \cite{Riotto}, is $n=2$. As this corresponds to a linear function, one can perform a field redefinition to bring  the induced Ansatz \eqref{induced} to \eqref{universal}. Therefore, at any value for the coupling, these two theories have to be identical. Indeed, taking a linear function for the universal Ansatz \eqref{universal} leads to a minimally coupling quadratic potential in the weak coupling limit. Therefore, taking the functions $f$ and $g$ to be linear leads to identical weak coupling limits. In all other cases, however, the weak coupling limits will be different while the strong coupling limits coincide.

One may also consider a more general set of polynomial functions $f  = \sum c_{n } \phi^{n}$. In the limit $\xi \to 0$, the position of the minimum of the potential appears at $\phi \to \infty$. Therefore for any particular choice of $c_{n}$ the predictions of the model will coincide with the prediction of the theory where $f(\phi) $ is dominated by the term $c_{n}\phi^{n}$ with the highest power $\phi^{n}$. This term determines the speed of the convergence to the predictions of the simplest model of chaotic inflation with a quadratic potential.

\smallskip

\noindent {\bf Discussion.\ } 
In this letter we have pointed out that the class of inflationary models that is referred to as induced inflation, as proposed by \cite{Giudice:2014toa}, has very interesting properties at strong and at weak coupling. In the first case, the entire class closely resembles Starobinsky inflation, with some subtle model-dependent differences. Equally surprising, the set of theories has a second attractor at weak coupling: in this opposite limit, the inflationary predictions converge to those of chaotic inflation with a quadratic potential. The proverbial ``opposites attract'' therefore very much applies to this class of inflationary theories. 

\smallskip

\noindent {\bf Acknowledgments.\ } 
RK and AL are supported by the SITP and by the NSF Grant PHY-1316699 and RK is also supported by the Templeton foundation grant `Quantum Gravity Frontiers'.

\end{document}